\documentclass[11pt]{article}

\usepackage{graphicx}
\usepackage{amsmath}
\usepackage[title]{appendix}
\usepackage[text={16cm,23cm},centering]{geometry}

\newcommand{\beq}{\begin{equation}}
\newcommand{\eeq}{\end{equation}}

\begin{document}

\title{Zipf's, Heaps' and Taylor's laws are determined by the expansion into the adjacent possible}

\author{%
Francesca Tria $^{1,*}$,
Vittorio Loreto $^{2,1,3}$, 
Vito~D.~P.~Servedio $^{3}$ \\[3mm]
{\footnotesize $^{1}$ Sapienza University of Rome, Physics Department, P.le Aldo Moro 5, 00185 Rome, Italy}\\
{\footnotesize $^{2}$ Sony Computer Science Laboratories, 6, rue Amyot, 75005, Paris, France}\\
{\footnotesize $^{3}$ Complexity Science Hub Vienna, Josefst\"adter Strasse 39, A-1080 Vienna, Austria}\\
{\footnotesize Correspondence: \texttt{$^*$ fratrig@gmail.com}}
}

\maketitle

\begin{abstract}
\noindent
Zipf's, Heaps' and Taylor's laws are ubiquitous in many different systems where innovation processes are at play. Together, they represent a compelling set of stylized facts regarding the overall statistics, the innovation rate and the scaling of fluctuations for systems as diverse as written texts and cities, ecological systems and stock markets. Many modeling schemes have been proposed in literature to explain those laws, but only recently a modeling framework has been introduced that accounts for the emergence of those laws without deducing the emergence of one of the laws from the others or without ad hoc assumptions.  This modeling framework is based on the concept of adjacent possible space and its key feature of being dynamically restructured while its boundaries get explored, i.e.,  conditional to the occurrence of novel events. Here, we illustrate this approach and show how this simple modelling framework, instantiated through a modified P\'olya's urn model, is able reproduce Zipf's, Heaps' and Taylor's laws within a unique self-consistent scheme. In addition the same modelling scheme embraces other less common evolutionary laws (Hoppe's model and Dirichlet processes) as particular cases.	
\end{abstract}
{\bf Keywords:} Innovation dynamics, Stylized facts, Zipf's law, Heaps' law, Taylor's law, Adjacent Possible, P\'olya's Urns, Poisson-Dirichlet processes.

\newpage
\tableofcontents
\newpage

\section{Introduction}
\noindent
Innovation processes are ubiquitous. New elements constantly appear in virtually all systems and the occurrence of the new goes well beyond what we now call innovation. The term innovation refers to a complex set of phenomena that includes not only the appearance of new elements  in a given system, e.g., technologies, ideas, words, cultural products, etc., but also their adoption by a given population of individuals. From this perspective one can distinguish between a personal, or local, experience of the new - for instance when we discover a new favorite writer or a new song - and a global occurrence of the new, i.e., every time something appears that never appeared before - for instance if we write a new book or write a new song. In all these cases there is something new entering the history of a given system or a given individual. 

Given the paramount relevance of innovation processes, it is highly important to grasp their nature and understand how the new emerges in all its possible instantiations. To this end, it is essential to fix a certain number of stylized facts characterizing the overall phenomenology of the new and quantifying its occurrence and its dynamical properties. Here we focus in particular on three basic laws whose general validity has been assessed in virtually all systems displaying innovation. The Zipf's law~\cite{estoup_1916,zipf_1929,zipf_1935,zipf_1949}, quantifying the frequency distribution of elements in a given system, the Heaps' law~\cite{herdan_1960,heaps_1978}, quantifying the rate at which new elements enter a given system and the Taylor's law~\cite{taylor_1961}, quantifying the intrinsic fluctuations of variables associated to the occurrence of the new. Any basic theory, supposedly close to the actual phenomenology of innovation processes, should be able at least to explain those three laws from first principles. Despite an abundant literature on the subject related to many different disciplines, a clear and self-consistent framework to explain the above-mentioned stylized facts, has been missing for a very long time. Many approaches have been proposed so far, often adopting ad-hoc assumptions or attempting to derive one of the three laws taking the others for granted. The aim of this paper is that of trying to put order in the often scattered and disordered literature, by proposing a self-consistent framework that, in its simplicity and generality, it is able to account for the existence of the three laws from very first principles. 

The framework we propose is based on the notion of "Adjacent Possible" and, more generally, on the interplay between what Francois Jacob named the dichotomy between the “actual” and the “possible”, the actual realization of a given phenomenon and the space of possibilities still unexplored. Originally introduced by the famous biologist and complex-systems scientist Stuart Kauffman, the  notion  of the adjacent possible~\cite{kauffman_1996,kauffman_2000} refers to the progressive expansion, or restucturing, of the space of possibilities, conditional to the occurrence of novel events. Based on this early intuition, we recently introduced, in collaboration with Steven Strogatz, a mathematical framework~\cite{adjacent_possible_2014,expanding_spaces_2016} to investigate the dynamics of the new via the adjacent possible. The modeling scheme is based on older schemes, named Polya’s urns and it mathematically predicts the notion that ``one thing leads to another'', i.e., the intuitive idea, presumably we all have, that innovation processes are non-linear and the conditions for the occurrence of a given event could realize only after something else happened. 

It turns out that the mathematical framework encoding the notion of adjacent possible  represents a sufficient first-principle scheme to explain the Zipf's, Heaps' and Taylor's laws on the same ground. In this paper we present this approach and we discuss the links it bears with other approaches. In particular we discuss the relation of our approach with well known stochastic processes, widely studied in the framework of nonparametric bayesian inference, namely the Dirichlet and the Poisson-Dirichlet processes~\cite{csp_pitman2006,buntine2010,deblasi2015}. Also based on this comparison, a coherent framework emerges where the importance of the adjacent possible scheme appears as crucial to understand the basic phenomenology of innovation processes. Though we can only conjecture that the expansion of the adjacent possible space is also a necessary condition for the validity of the three laws mentioned above, no counterexamples have been found so far that, without a dynamical space of possibilities, one can satisfactorily explain the empirically observed laws. 

\section{Zipf's and Heaps' laws}

\subsection{Frequency-rank relations: the Estoup-Zipf's law}

\begin{figure}[t]
\centering
\includegraphics[width=0.6\textwidth]{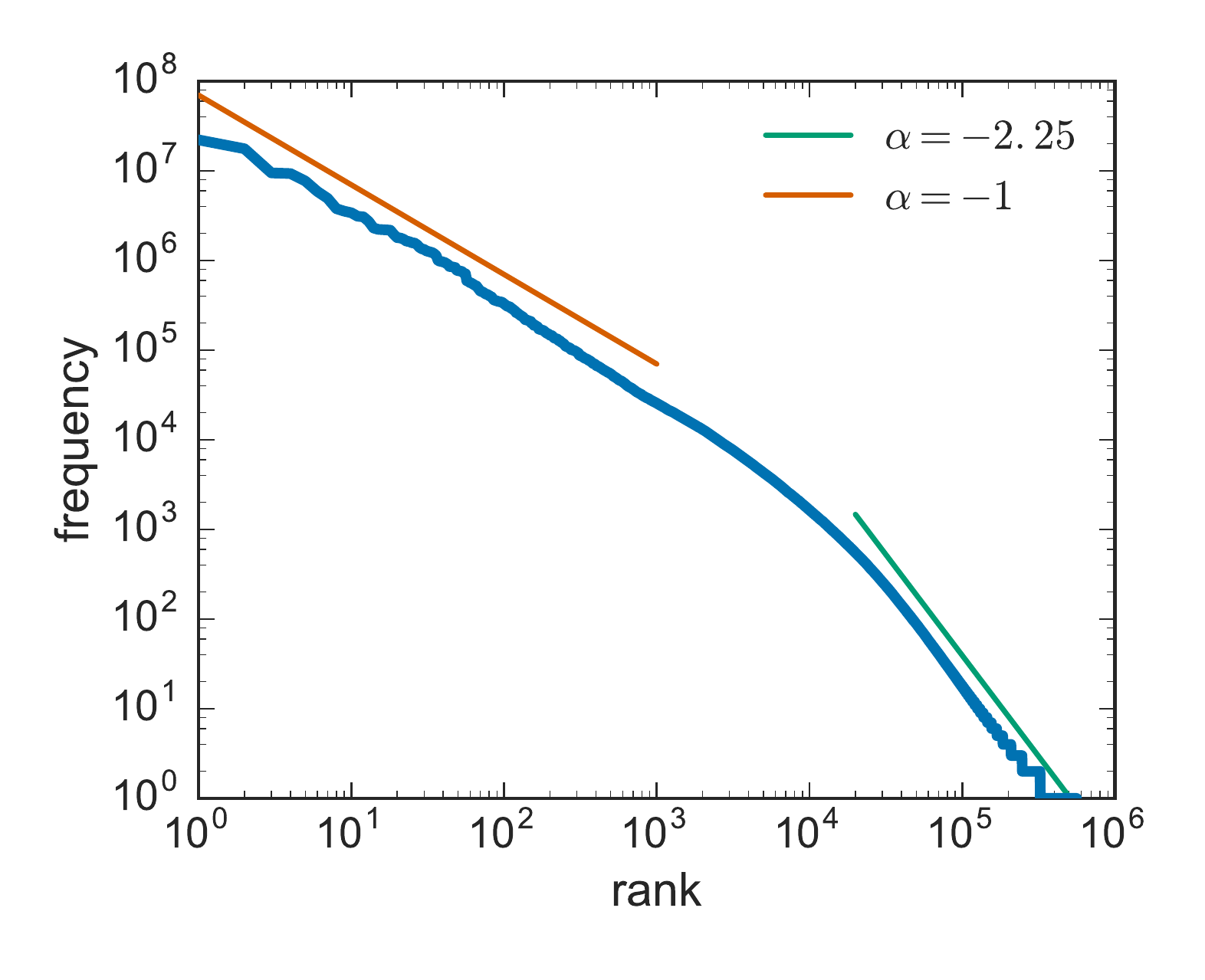}
\caption{{\bf Zipf's law} computed on the Gutenberg corpus~\protect\cite{gutenberg}. 
In this case the exponent of the asymptotic behaviour  is $\alpha \simeq 2.25$. Similar behaviours are observed in many other systems.}
\label{fig:gutenberg-zipf}
\end{figure}

\noindent Let's consider a generic text and count the number of occurrences of each word. Now, suppose to repeat the same operation for all the distinct words in a long text, rank all the words according to their frequency of occurrence and plot in a graph the number of occurrences vs. the rank. This is what George Kingsley Zipf did~\cite{zipf_1929,zipf_1935,zipf_1949} in the twenties of the XX century. A more recent analysis of the same behaviour is reported in Fig.~\ref{fig:gutenberg-zipf}, based on data of the Gutenberg corpus~\cite{gutenberg}.

\noindent  The existence of straight lines in the log-log plot is the signature of power-law functions of the form:
\begin{equation}
f(R) \sim R^{-\alpha}
\end{equation}
\noindent The original result obtained by Zipf, corresponding to the first slope with $\alpha \simeq 1$, revealed a striking regularity in the way words are adopted in texts: said $f(1)$ the frequency of the most frequent word (rank $R=1$), the frequency of the second most frequent word is $f(1)/2$, that of the third $f(1)/3$ and so on. For high rankings, i.e., highly infrequent words, one observes a second slope, with an exponent larger than two.  

It should be remarked that perhaps the first one to observe the above reported law was Jean-Baptiste Estoup, who was the General Secretary of the Institut St\'enographique de France. In his book {\em Gammes st\'enographiques}~\cite{estoup_1916,petruszewycz_1973}, pioneered the investigation of the regularity of word frequencies and observed that the frequency with which words are used in texts appears to follow a power law behaviour. This observation was later acknowledged by Zipf~\cite{zipf_1929} and examined in depth to bring to what is also known as the Estoup-Zipf's law. From now onwards we shall refer to this law as the Zipf's law.

It is also important to remark that Zipf-like behaviours have been observed in a large variety of cases and situations. Zipf itself reported~\cite{zipf_1949} about the distribution of  metropolitan districts in 1940 in the US, service establishments, manufacturers, retails stores in the USA in 1939. Along the years the number of examples and situations where the Zipf's law has been invoked ha been steadily growing:  for instance cities populations, the statistics of webpage visits and other internet traffic data, company sizes and other economic data, science citations and other bibliometric data, scaling in natural and physical phenomena. A thorough account of all these cases is out of the scope of the present paper and we refer to recent reviews and references therein for an account of the latest developments~\cite{Li_2002,newman_2005,Piantadosi_2014}.

\subsection{The innovation rate: the Herdan-Heaps' law}

\begin{figure}[t]
\centering
\includegraphics[width=0.6\textwidth]{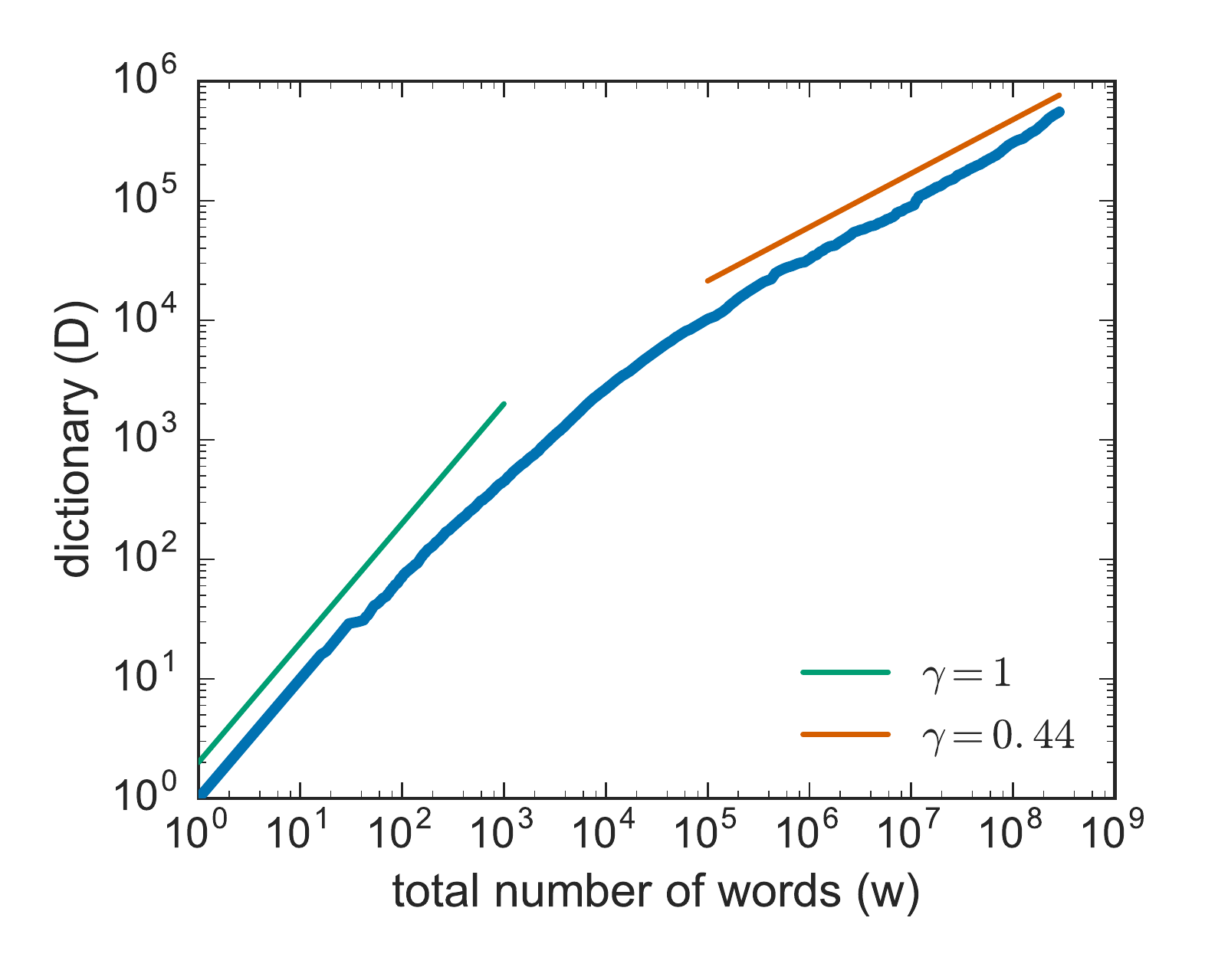}
\caption{Growth of the number of distinct words computed on the Gutenberg corpus of texts~\protect\cite{gutenberg}. The position of texts in the corpus is chosen at random. In this case $\gamma \simeq 0.44$. Similar behaviours are observed in many other systems.}
\label{fig:gutenberg-heaps}
\end{figure}

Let us now make a step forward and look at a generic text (or, without loss of generality, at a generic sequence of characters) and focus now on the occurrence of the novelties. For a generic text one can ask when new words, i.e., never occurred before in the text, appear. Now, if one plots the number of new words as a function of the number of words read (which is our measure of the intrinsic time), one gets a plot like that of Fig.~(\ref{fig:gutenberg-heaps}), where one observes two main behaviors.
\noindent  A linear growth for short times where basically at the beginning all the words are appearing for the first time. Later on the growth slows down and an asymptotic behavior it is observed of the form:
\begin{equation}
\label{eq:heaps}
D(N) \sim N^{\gamma} 
\end{equation}

\noindent with $\gamma \in [0,1]$. In the specific case of Fig.~(\ref{fig:gutenberg-heaps}) $\gamma \simeq 0.45$ but the exponent slightly changes from text to text. The relation of Eq.~(\ref{eq:heaps}) is known as Heaps' law from Harold Stanley Heaps~\cite{heaps_1978}, who formulated it in the framework of information retrieval (see also~\cite{Baeza-Yates_2000}), though its first discovery is due to Gustav Herdan~\cite{herdan_1960} in the framework of linguistics (see also~\cite{baayen_2001,Egghe_2007}).  From now onward we shall refer to it as Heaps' law.

\subsection{Zipf's vs. Heaps' laws}\label{sec:IntroTaylor} 
\label{sec:zipf_vs_heaps}
\noindent
In this section we compare the two laws just observed, the Zipf's law for the frequencies of occurrence of the elements in a system and the Heaps' law for their temporal appearance. It has been often claimed that Heaps' and Zipf's law are trivially related and that one can derive the Heaps's law once the Zipf's is known.This is not true in general.  It turns out to be true only under the specific hypothesis of random-sampling as follows. 
Suppose the existence of a strict power-law behaviour of the frequency-rank distribution, $f(R)\sim R^{-\alpha}$, and construct a sequence of elements by randomly sampling from this Zipf distribution $f(R)$. Through this procedure one recovers a Heaps' law with the functional form $D(t) \sim t^{\gamma}$~\cite{serrano_2009,lu_2010} with $\gamma=1/\alpha$.
In order to do that we need to consider the correct expression for $f(R)$ that includes the normalisation factor, whose expression can be derived through the following approximated integral:
\begin{equation}
\int_{1}^{R_{\max}} f(\tilde{R}) d  \tilde{R} = 1 \,.
\end{equation}

\noindent Let us now distinguish the two cases. For $\alpha \neq 1$ one has 
\begin{equation}
\label{eq:zipf}
f(R) =  \frac {1-\alpha}{R_{\max}^{1-\alpha} -1}  R^{-\alpha} \,.
\end{equation}

\noindent while for $\alpha = 1$ one obtains:
\begin{equation}
f(R) = \frac {1} {\log{R_{\max}}} R^{-1} \,.
\end{equation}

\noindent When $\alpha >1$, one can neglect the term $R_{\max}^{1-\alpha}$ in Eq.(\ref{eq:zipf}), and when $\alpha <1$, one can write $R_{\max}^{1-\alpha} -1 \simeq R_{\max}^{1-\alpha} $.\\
Summarizing one has then:

\begin{eqnarray}
\alpha >1 ~ : && ~~~ f(R) \simeq  (\alpha -1)  R^{-\alpha} \,. \label{eqn1:zipf}\\
&& \nonumber \\
\alpha = 1  ~ : && ~~~ f(R) \simeq \frac {R^{-1}} {\ln{R_{\max}}} \,. \label{eqn2:zipf}\\
&& \nonumber \\
0<\alpha<1 ~  : && ~~~ f(R) \simeq (1-\alpha) \frac {R^{-\alpha}}{R_{\max}^{1-\alpha} }  \,.
\label{eqn3:zipf}
\end{eqnarray}

We are now interested in estimating the number, $D$, of distinct elements appearing in the sequence as a function of its length $N$. To do that, let us consider the entrance of a new element (never appeared before) in the sequence and let the number of distinct elements in the sequence be $D$  after this entrance. This new element will have maximum rank $R_{\max} =D$, and frequency $f( R_{\max}) = 1/N$.  From Eq.~(\ref{eqn1:zipf}) we obtain:
\begin{equation}
f(D) \simeq  (\alpha -1)  D^{-\alpha}  = \frac{1}{N}\,
\end{equation}

\noindent which, after an inversion gives:
\begin{eqnarray}
D \simeq N^{\gamma} \,\,\,\,\,\,\,\,\,\,\, \text{with} \,\,\,\,\,\,\,\,\,\,\,  && \gamma= \frac{1}{\alpha}
\end{eqnarray}

\noindent The same reasoning can be extended to generic functional forms for $D(N)$ as follows:
\begin{eqnarray}
\alpha >1 \,\,\,: && \,\,\,\,\,\,
f(D) \simeq  (\alpha -1)  D^{-\alpha}  = \frac{1}{N}\,. \\
&& \nonumber \\
\alpha = 1  \,\,\,: &&\,\,\,\,\,\,
 f(D) \simeq \frac {1} {D \ln{D}} = \frac{1}{N}\,. \\
&& \nonumber \\
0<\alpha<1\,\,\,: && \,\,\,\,\,\, 
 f(D) \simeq \frac {1-\alpha}{D^{1-\alpha} -1}  D^{-\alpha} = \frac{1}{N} \,.
\end{eqnarray}

Inverting these relations one eventually finds:
\begin{eqnarray}
\alpha >1\ : && ~~
D \simeq N^{\gamma} ~~~ \text{with} ~~~  \gamma= 1/\alpha \label{eqn1:heaps}\\
\alpha = 1\ : && ~~
 D \simeq N/\ln N ~~~ \text{with} ~~~  \gamma \simeq 1 \label{eqn2:heaps}\\
0<\alpha<1\ : && ~~ 
 D \simeq N ~~~ \text{with} ~~~  \gamma= 1\,. \label{eqn3:heaps}
\end{eqnarray}

\noindent Summarizing, under the hypothesis of random sampling from a frequency rank distribution expressed by a power-law function $f(R)\sim R^{-\alpha}$, one recovers a Heaps' law $D(N) \sim N^{\gamma}$ with the following relation between $\gamma$ and $\alpha$:
\begin{eqnarray}
\alpha >1\ : && ~~ \gamma=1/{\alpha} \nonumber \\
0<\alpha \leq 1\ : && ~~~ \gamma =1.
\label{eq:heaps_zipf_finite_size}
\end{eqnarray}
In~\cite{lu_2010} it has been demonstrated that finite-size effects can affect the above-seen relationships that happen to be true only for very long sequences. For short enough sequences one observes a systematic deviation from Eq.~(\ref{eq:heaps_zipf_finite_size}), especially for $\alpha$ values close to $1$.
\newline
Another important observation is now in order. The assumption of random sampling considered above is strong and sometimes unrealistic (e.g.,~\cite{Cristelli_2012}). First of all it implies the \emph{a priori} existence of a Zipf's law with an infinite support. In addition, the frequency-rank plots one empirically observes are far from featuring a pure power-law behavior. In all those cases the relation between the Zipf's law and the Heaps' law seen above and summarized by  Eq.~(\ref{eq:heaps_zipf_finite_size}) happens to hold only when looking at the tail of the {Z}ipf's plot, i.e., for high ranks (small frequencies) in the frequency-rank plots and long times, i.e., high $N$, in the plot expressing the Heaps' law. In a later section we shall also discuss the so-called Taylor's law that connects the standard deviation $s$ of a random variable (for instance the size $D$ of the dictionary) to its mean $\mu$. Simple analytic calculations~\cite{gerlach2014} show that the poissonian sampling of a power-law leads to a Taylor's law with exponent $1/2$, i.e., $s\propto \sqrt\mu$.  This is not the case for real texts for which one observes an exponent close to $1$~\cite{gerlach2014}. 

The ensemble of all these facts imply that the explanation of the empirical findings of the Zipf's and Heaps' law cannot be done by only deriving one of the law and deducing the other one accordingly, based on Eq.~(\ref{eq:heaps_zipf_finite_size}). Rather both Zipf's and Heaps' laws and the Taylor's law should be all derived in the framework of a self-consistent theory. This is precisely the aim of this paper.

\section{Urn model with triggering}
\noindent
We now introduce a simple modeling scheme able to reproduce both Zipfs and Heaps' laws simultaneously. Crucial for this result is  the conditional  expansion of the space of possibilities, that we will elucidate in the following. In~\cite{kauffman_1996,kauffman_2000} S. Kauffman introduces and discusses the notion of the {\em adjacent possible}, that is of all those things that are one step away from what actually exists. The idea is that evolution does not proceed by leaps, but moves in a space where each element should be connected with its precursor. The Kauffman's theoretical concept of  {\em adjacent possible},  originally discussed in his investigations of molecular and biological evolution,  has also been applied to the study of innovation and technological evolution~\cite{johnson_2010_book,wagner2014spaces}. To clarify the concept, let us think to a baby that is learning to talk.  We can  say almost  surely that she will not utter "serendipity" as the first word in her life. More than this, we can safely guess that her first word will be "papa", or "mama", or one among a  list of  few other possibilities. In other words, in the period of lallation only few words belong to the space of the adjacent possible and can be actualized  in the next future. Once the baby has learned how to utter simple words, she can try more sophisticated ones, involving more demanding articulation efforts. In the process of learning, her space of possibilities (her adjacent possible) considerably grows, with the result that guessing a priori the hundredth words learned by a child is much less obvious than guessing which will be the first one. 

Here we formalize this idea that by opening up new possibilities, an innovation paves the way for other innovations, explicitly introducing this concept in a  P\'olya's urn based model. In particular, we will discuss  the simplest version of the model introduced in~\cite{adjacent_possible_2014}, that we will name  P\'olya's urn model with triggering (PUT). The interest of this model lies, on the one hand, in its generality, the only assumptions it makes refer to the general and not system-specific mechanisms for the expansion into the adjacent possible; on the other hand its simplicity allows to draw analytical solutions. 

\begin{figure}[t]
\centerline{\includegraphics[width=\textwidth]{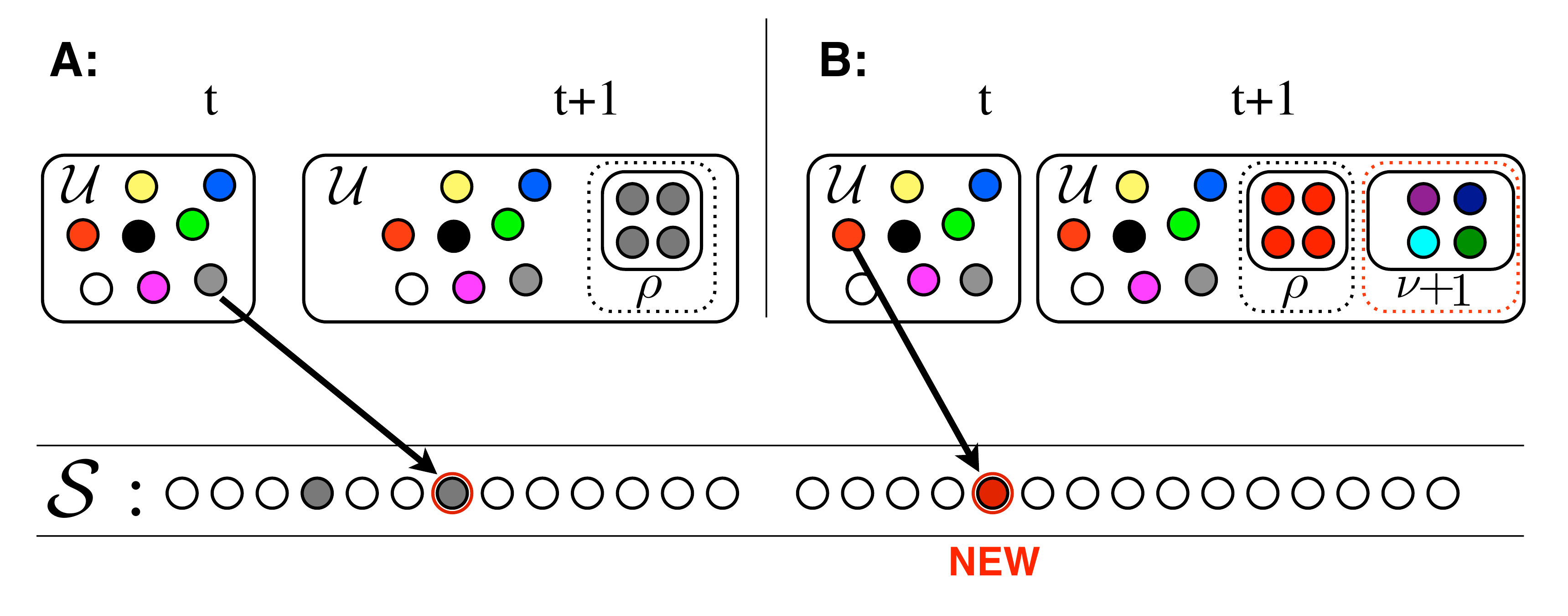}}
\caption{ {\bf Urn model with triggering.} {\bf A:}
 An element that had previously been drawn from the urn, is drawn again: the element is added to $\mathcal{S}$ and it is put back in the urn along with $\rho$ additional copies of it. {\bf B:} An element that  never appeared in the sequence is drawn: the element is added to $\mathcal{S}$, put back in the urn along with $\rho$ additional copies of it, and $\nu+1$ brand new and distinct balls are also added to the urn.}
\label{fig:model}
\end{figure}

The model works as follows (please refer to Fig.~\ref{fig:model}). An urn $\mathcal{U}$ initially contains $N_0$ distinct elements, represented by balls of different colors. By randomly extracting elements from the urn, we construct a  sequence $\mathcal{S}$ mimicking the evolution of our system (e.g., the sequence of words in a given text). Both the urn and the sequence enlarge during the process: (i) at each time step $t$, an element $s_t$ is drawn at random from the urn, added to the sequence, and put back in the urn along with $\rho$ additional copies of it (Fig.~\ref{fig:model},  A); (ii) iff the chosen element $s_t$ is new (i.e., it appears for the {\em first time} in the sequence $\mathcal{S}$), $\nu+1$ brand new distinct elements are also added to the urn (Fig.~\ref{fig:model}, B). These new elements represent the set of new possibilities opened up by the seed $s_t$. Hence $\nu+1$ is the size of the {\em new adjacent possible} available once an innovation occurs.

\begin{figure}[t]
\centerline{\includegraphics[width=\textwidth]{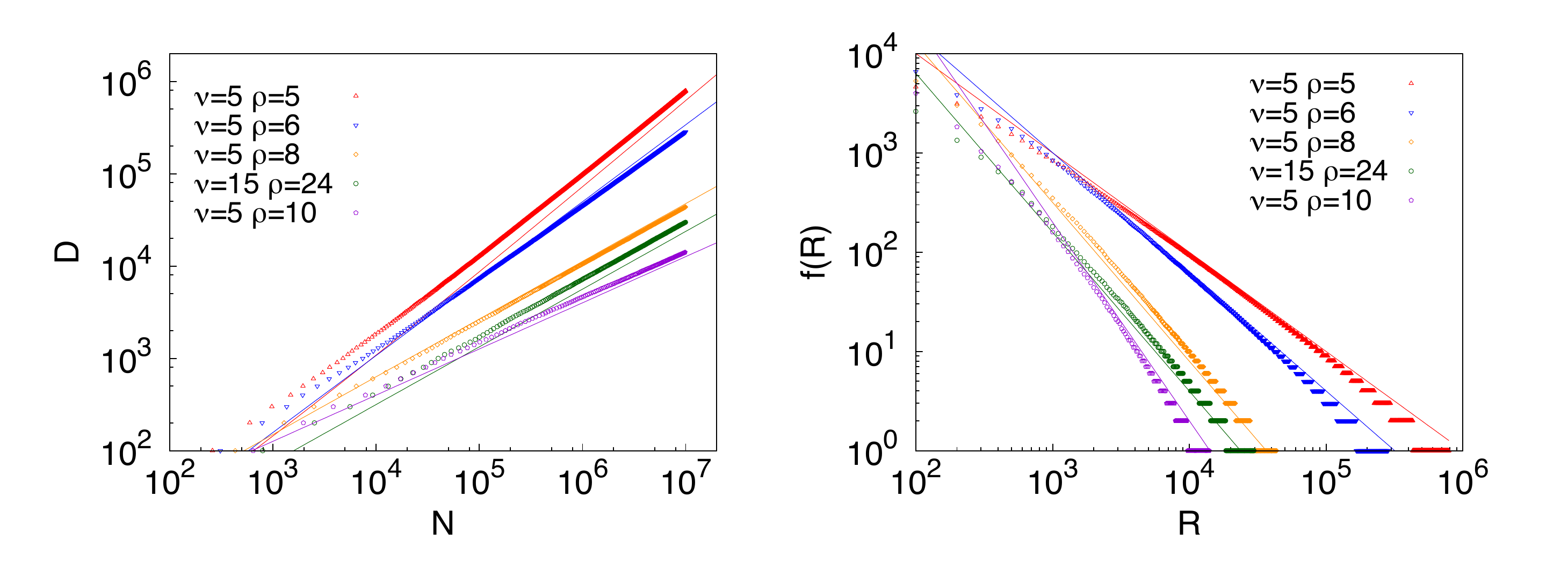}}
\caption{ {\bf Heaps' law (left) and Zipf's law (right) in  the urn model with triggering.}  Straight lines in the Heaps' law plots show functions of the form $f(x)=a x^{\gamma}$ with the exponent $\gamma=\nu/\rho$ as predicted by the analytic results and confirmed in the numerical simulations. Straight lines in the Zipf's law plots show functions of the form $f(x)=a x^{-\alpha}$, with $\alpha = \gamma^{-1}=\rho/\nu$.  }
\label{fig:heaps_zipf_model}
\end{figure}

\noindent Simple asymptotic formulas for the number $D(n)$ of distinct elements appearing in the sequence as a function of the sequence's length $n$ (Heaps' law), and for the asymptotic power-law behavior of the frequency-rank distribution (Zipf's law),  in terms of  the model parameters $\rho$ and $\nu$ can be derived. In order to do so, one can write a recursive formula for $D(n)$ as:
\begin{equation}
D(n+1)=D(n)+ P_N(n)
\label{eq:discr}
\end{equation}
where we have defined $P_N(n)$ as the probability of drawing a  new ball (never extracted before) at time $n$ (note that we consider intrinsic time, that is we identify the time elapsed with the length of the sequence constructed). The probability $P_N(n)$ is equal to the ratio (at time $n$) between the number of elements in the urn never extracted and the total number of elements  in the urn. Approximating eq.~(\ref{eq:discr}) with its continuous limit, we can write:
\begin{equation}
  \frac{d D}{dn}= \frac{N_0 + \nu D}{N_0+(\nu+1) D+\rho n},\label{eq:D_complete_text}
\end{equation}
where $N_0$ is the number of balls, all distinct, initially placed in the urn. This equation can be integrated analytically in the limit of large $n$, when $N_0$ can be neglected, by performing a change of variable $z=\frac{D}{n}$. After some algebra (detailed computations can be found in~\cite{expanding_spaces_2016} and in Appendix A for an extended model), we find the 
 asymptotic solutions (valid for large $n$):
\begin{eqnarray}
 \rho > \nu & \Rightarrow & D (n) \sim   \left(\frac{\rho-\nu}{\rho+1}\right)^{\frac{\nu}{\rho}}  n^{\frac{\nu}{\rho}} \,;\label{eq:asymptotic_sub}\\
 \rho < \nu & \Rightarrow & D (n)  \sim \frac{\nu-\rho}{\nu+1}n \, ;\label{eq:asymptotic_lin}\\
 \rho = \nu & \Rightarrow & D (n)  \log D \sim \frac{\nu}{\nu+1} n \rightarrow D \sim \frac{\nu}{\nu+1} \frac{n}{\log n} \, \label{eq:asymptotic_log}.
\end{eqnarray}
For the derivation of the Zipf's law we refer the reader to the SI of~\cite{adjacent_possible_2014} and to Appendix~\ref{sec:appendixB} for an alternative derivation based on the continuous approximation.
Results contrasting numerical results and theoretical predictions for the Heaps' and Zipf's laws are reported in Fig.\ref{fig:heaps_zipf_model}.

\subsection{The role of the adjacent possible: Heaps' and Zipf's laws in the classic multicolors P\'olya urn model}
\begin{figure}
\centerline{\includegraphics[width=0.45\textwidth]{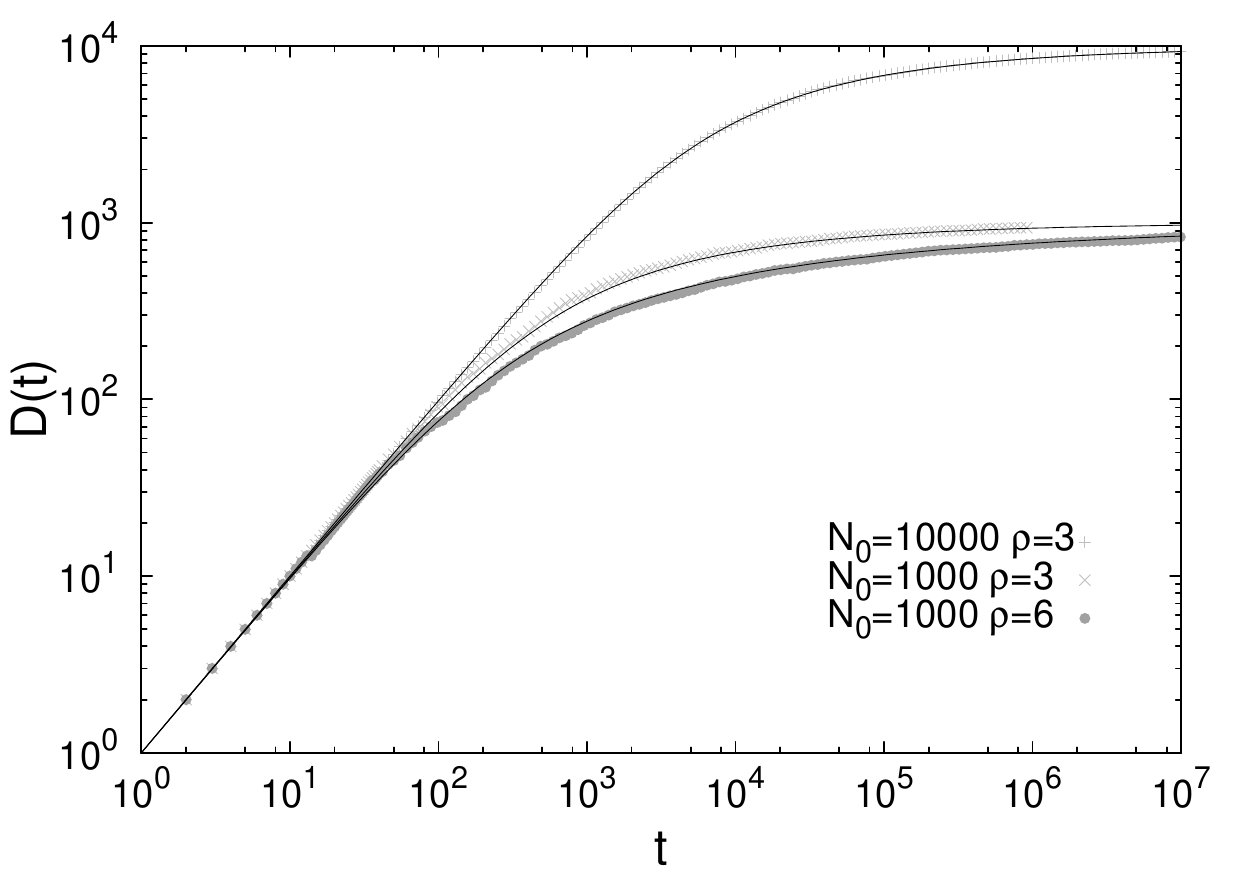}%
\includegraphics[width=0.45\textwidth]{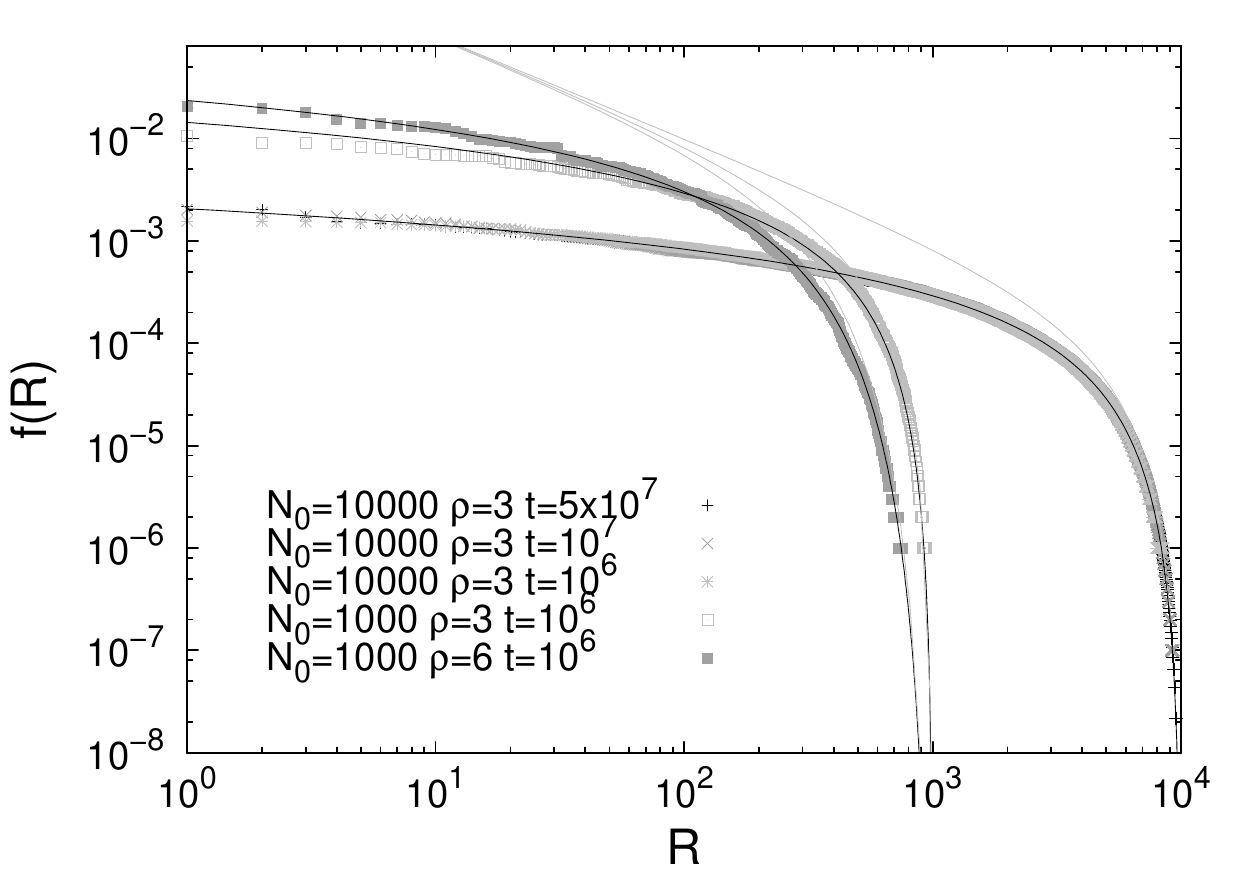}}
\caption{{\bf Results for the multicolors P\'olya urn model without innovation.} Results  are reported both from simulations of the process (points) and from the analytical  predictions (straight lines), for different values of the initial number  of balls $N_0$ and of the reinforcement parameter $\rho$.  LEFT: Number of different colors $D(n)$ added in the sequence as a function of the total number $t$ of extracted balls.  The curves from analytical predictions of Eq.~(\ref{eq:Dt_polya_multicolors}) exactly overlap the simulated points. RIGHT: Frequency-rank distribution. Simulations of the process are here reported along with both: (i) the prediction obtained by inverting the relation $R \simeq = \frac{N_0}{\Gamma \left(\frac{1}{\rho}\right) }\Gamma\left( \frac{1}{\rho}, \frac{N_0-1 -\rho}{\rho} f\right)$; (ii) the asymptotic solution, valid for $R\gg1$, obtained by inverting Eq.~(\ref{eq:Dt_polya_multicolors}) $f(R)\simeq \frac{\rho}{N_0}\left[\left(1-\frac{R}{N_0}\right)^{-r} -1\right]^{-1}$ (refer to~\cite{Tria_cimento2016} for their derivation).}
\label{fig:polya_multicolors}
\end{figure}

\noindent
One question that  naturally emerges concerns the relevance of the notion of adjacent possible and its conditional growth. One could for instance argue that the same predictions of the PUT model could be replicated having all the possible outcomes of a process immediately available from the outset, instead of appearing progressively through the conditional process related to the very notion of adjacent possible. In order to remove all doubt, we consider an urn initially filled with $N_0$ distinct colors, with $N_0$ arbitrarily large, with no other colors entering in the urn during the process of construction of the sequence $\mathcal{S}$. This is the  P\'olya multicolors urn model~\cite{polya_multicolor_1997} and  we here briefly discuss the Heaps' and Zipf's laws  emerging from it.  Let us thus consider an urn initially containing $N_0$ balls, all of different colors. At each time step a  ball is withdrawn at random, added to a sequence, and placed back in the urn along with $\rho$ additional copies of it. This process corresponds to the one depicted in Fig.~(\ref{fig:model}) A, that is to the rule of the PUT model in the case the drawn element is not new.
\newline
Note that although in this case the urn does not acquire new colors during the process,  we can still study the dynamic of innovation  by looking at the entrance of new color in the growing sequence. Let us then consider $N_0$ very large, so that we can consider a long time interval far from saturation (when there are still many colors in the urn not already appeared in $\mathcal{S}$). The number of different colors $D(n)$ added to the sequence at time $n$ follows the equation (when the continuous limit is taken):
\begin{equation}\label{eq:Dt_polya_multicolors}
\frac{dD}{dn}=\frac{N_0-D(n)}{N_0+\rho n}  \, , \,\,\,\,\,\, D(0)=0
\,\,\,\,\,\,   \Rightarrow \,\,\,\,\,\,  D(n) =N_0\left[1-\left( 1 +\frac{\rho n}{N_0}\right)^{
    -\frac{1}{\rho}}\right]   \,.
\end{equation}
We thus obtain  that for $\rho n \ll N_0$, $D(n)$ follows a linear behaviour  ($D(n)  \simeq n$), while for large $n$ saturates at $D(n) \simeq N_0$, failing to predict the power law (sublinear) growth of new elements. In figure~\ref{fig:polya_multicolors} we report results for both the Heaps' and Zipf's laws predicted by the model along with their theoretical predictions, referring the reader to~\cite{Tria_cimento2016} for a detailed derivation of the Zipf's law. It is evident how a simple exploration of a static, though large, space of possibilities, cannot account for the empirical observations summarized by the Zipf's and the Heaps's laws.

\section{Connection of the urn model with triggering with stochastic processes featuring innovation}
\noindent
The PUT model is closely related to well known stochastic processes, widely studied in the framework of nonparametric bayesian inference, namely the Dirichlet and the Poisson-Dirichlet processes. We will discuss here those processes in term of their predictive probabilities, referring  to excellent reviews~\cite{csp_pitman2006,buntine2010, deblasi2015}  for a complete and formal definition of them.
 
The problem can be framed in the following way. Given a sequence of events $x_1, \cdots , x_n$, we want to estimate the probability that the next event will be $\tilde{x}$, where $\tilde{x}$ can be one of the already seen events $x_i$, $i=1, \cdots , n$, or a completely new one, unseen till the intrinsic time $n$.

\subsection{Urn model with triggering and the Poisson-Dirichlet process}

Let us first discuss the   Poisson-Dirichlet process, whose predictive conditional probability reads:
 \begin{equation}\label{eq:conditional_PD}
  p(\tilde{x} | x_1, \cdots , x_n; \alpha, \theta, p_0 ) = \frac{\theta + D \alpha}{\theta+n} P_0(\tilde{x}) + \sum_{i}^{D} \frac{n_i-\alpha}{\theta+n} \delta_{\tilde{x}_i,\tilde{x}}
 \end{equation}
where $0\leq\alpha<1$ and $\theta> - \alpha$ are parameters of the model, $P_0$ a given  continuous probability distribution defined {\em a priori} on the possible values the variables $x_i$ can take, named {\em base probability distribution},
and $\tilde{x}_i$ the $D$ distinct values appearing  in the sequence $x_1, \cdots , x_n$, respectively with multiplicity $n_i$. Let us briefly discuss Eq.~(\ref{eq:conditional_PD}). The first term in the  right hand side refers to the probability that $x_{n+1}$ takes a value never appeared before, i.e., a novel event. This happens with  probability $\frac{\theta + D \alpha}{\theta+n}$, depending both on the total number $n$ of events seen till time $n$, and on the  total number $D$ of distinct events seen till time $n$. In this way, in the Poisson-Dirichlet process it is implicit the concept that more novelties are actualized, higher is the probability of encountering further novelties. The second term in Eq.~(\ref{eq:conditional_PD}) weights the probability that $x_{n+1}$ equals one of the previously occurred events, and differs from a bare proportionality rules when $\alpha>0$.

The   Poisson-Dirichlet process predicts an asymptotic power-law behavior for the number $D(n)$ of distinct elements seen as a function of the sequence length $n$. The exact expression for the expected value of $D(n)$ can be found in~\cite{csp_pitman2006}. Here we report the results obtained under the same approximations made for the urn model with triggering: 
\begin{equation}
  \frac{d D}{dn}= \frac{\theta + \alpha D}{\theta+ n} ~,  ~~~ D(0)=0 \,,\label{eq:D_approx_PY}
\end{equation}
that can be solved by separation of variables, leading to:
\begin{equation}
  D (n) \sim \frac{\theta^{1- \alpha } (\theta+n)^\alpha}{\alpha} -\frac{\theta}{\alpha} ~. \label{eq:D_sol_PY}
\end{equation}

\noindent Note that the Poisson-Dirichlet process predicts a sublinear power law behavior for $D(n)$ but cannot reproduce a linear growth for it, being only defined for $\alpha <1$.

The ubiquity of the Poisson-Dirichlet process is due, together with its ability of producing sequences featuring Heaps' and Zipf's laws,  to the fundamental property of \emph{exchangeability}~\cite{Pitman1995, csp_pitman2006}. This refers to the fact that the probability of a sequence generated by the Poisson-Dirichlet process does not depend on the order of the elements in the sequence: $p( x_1, \cdots , x_n; \alpha, \theta, p_0 ) = p( \pi(x_1), \cdots , \pi(x_n); \alpha, \theta, p_0 )$ for any permutation $\pi$ of the sequence elements, so that we can  write the joint probability distribution $p( n_1, \cdots , n_D; \alpha, \theta, p_0 ) $ for the number of occurrences of the variables $x_i$.  Exchangeability is a powerful property, related to the de Finetti theorem~\cite{deFinetti1937,Zabell_1992}, nevertheless is also a  strong and sometimes unrealistic assumption on the lack of correlations and causality in the data.

Coming back to the PUT model, we observe that the model produces in general sequences that are not exchangeable. It recovers exchangeability in a particular case, corresponding to a slightly different definition of rule (i):  the drawn element $s_t$ is  put back in the urn along with $\rho$ additional copies of it iff $s_t$ is not new; in the other case (i.e., when we apply rule (ii)), $s_t$ is  put back in the urn along with  $\tilde{\rho}$ additional copies of it, with $\tilde{\rho}=\rho-(\nu+1)$. In this particular case the PUT model corresponds exactly to the Poisson-Dirichlet process, with $\theta=\frac{N_0}{\rho}$ and $\alpha=\frac{\nu}{\rho}$. In this case, at odds with the previously discussed version of the model,  the urn acquires  the same number of balls at each time step, regardless whether a novelty occurs or not.  This variant makes the generated sequences exchangeable, but imposes the constraint $\rho \geq (\nu+1)$, and thus in this case we cannot recover the linear growth of $D(n)$, as well as in the Poisson-Dirichlet process.  We demonstrate in Appendix A that the dependence of the power law's exponents  of the Heaps' and Zipf's laws on the PUT model's parameters $\rho $ and $\nu$ reads the same  as in Eq.s~(\ref{eq:asymptotic_sub})-(\ref{eq:asymptotic_log})  if we modify rule (i) with any $\tilde{\rho} \geq 0$.
  
Here we wish to remark that the urn representation of the PUT model allows for straightforward generalizations where correlations can be explicitly taken into account (see for instance~\cite{adjacent_possible_2014} for a first step in this direction). In addition it can be easily rephrased in terms of  walks in a complex space (for instance a graph), allowing to consider more complex underlying structures for the space of possibilities (see for instance the SI of~\cite{adjacent_possible_2014},~\cite{Monechi2016} and~\cite{Latora2018}). 

\subsection{Urn model with triggering, Dirichlet process and Hoppe model}
\noindent 
By setting $\alpha=0$ in Eq.~(\ref{eq:conditional_PD}), we obtain the predictive conditional probability for the Dirichlet process, predicting a logarithmic growth of $D(n)$~\cite{csp_pitman2006}. Correspondingly, if we chose $\nu=0$ in the urn model, we obtain:   
\begin{equation}
  \frac{d D}{dn}= \frac{N_0 }{N_0+ D+\rho n}
  \label{eq:D_log_urn}
\end{equation}
If we now neglect $D(n)$ in the denominator of~(\ref{eq:D_log_urn}), we can solve in the large limit of large $n$:
\begin{equation}
  D(n) \sim \frac{N_0 }{\rho} \log{\left(1+ \frac{ \rho}{N_0} n\right)}.\label{eq:D_log_urn_sol}
\end{equation}

The same asymptotic growth of $D(n)$ is also found in one of the first model introducing innovation in the framework of P\'olya's urn, namely the  Hoppe's model~\cite{Hoppe_1984}. The motivation of the Hoppe's work was to derive  the Ewens' sampling formula~\cite{Ewens_1972},  describing the allelic partition at equilibrium of a sample from a population evolved according to a discrete time Wright-Fisher process~\cite{fisher1930,wright1931evolution}. In the Hoppe model innovations are introduced through a special ball, the ``mutator''. In particular, the process starts with only the mutator in the urn, with a mass $\theta$.   At any time $n$, a ball is withdrawn with a probability proportional to its mass, and, if the ball is the mutator, it is placed back in the urn along with a ball of a brand new color, with unitary mass, thus increasing the number of different colors present in the urn.  Otherwise, the selected ball is placed back in the urn along with  another ball of the same color.  Writing the recursive formula for $D(n)$ and taking the continuous limit, we obtain:
\begin{equation}
\frac{d D(n)}{  dn } = \frac{\theta}{\theta+n} ~,  ~~~ D(0)=0 \,,
\end{equation}
that is exactly Eq.~(\ref{eq:D_log_urn}) with $\alpha=0$. It predicts a logarithmic increase of new colors in the urn:
\begin{equation}\label{eq:Dt_hoppe}
D(n) = \theta \ln {(\theta +n)} - \theta \ln {(\theta)}  = 
    \theta\ln (1+\frac{n}{\theta})\,,
\end{equation}
corresponds to Eq.~(\ref{eq:D_log_urn_sol}), by identifying $\frac{N_0}{\rho}$ with $\theta$. Hoppe's urn scheme is non-cooperative in the sense that one novelty does not nothing to facilitate another. In other words, while in the Hoppe model it is already present a mechanism that allows for the expansion of the space of possibilities, this mechanism is completely independent  on the actual realization of a novelty, and fails to reproduce  both the Heaps' and the Zipf's laws.

\section{Fluctuation scaling (Taylor's law)}
\label{sec:taylor}
\begin{figure}[t]
\centerline{\includegraphics[width=0.45\textwidth]{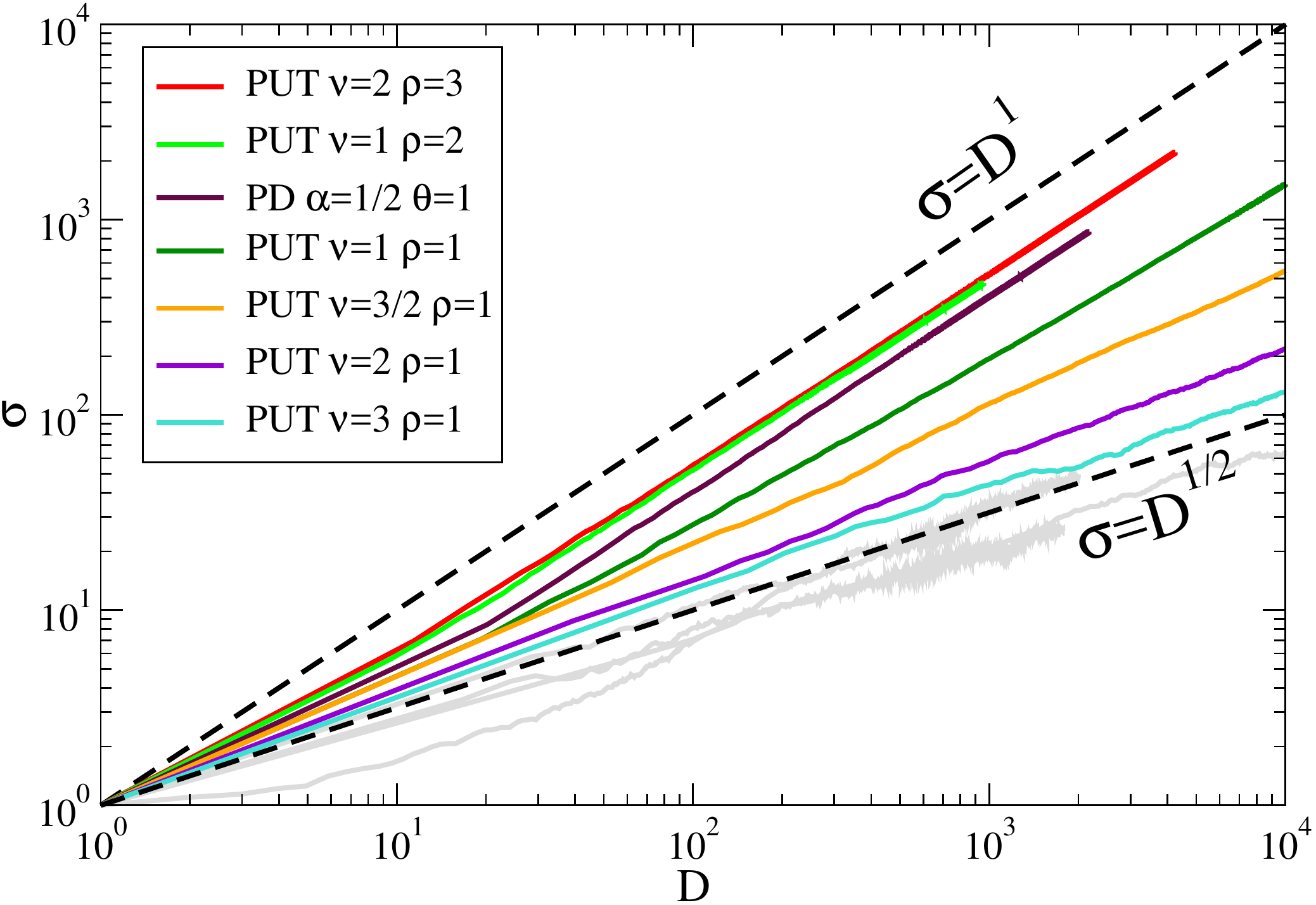}%
\includegraphics[width=0.45\textwidth]{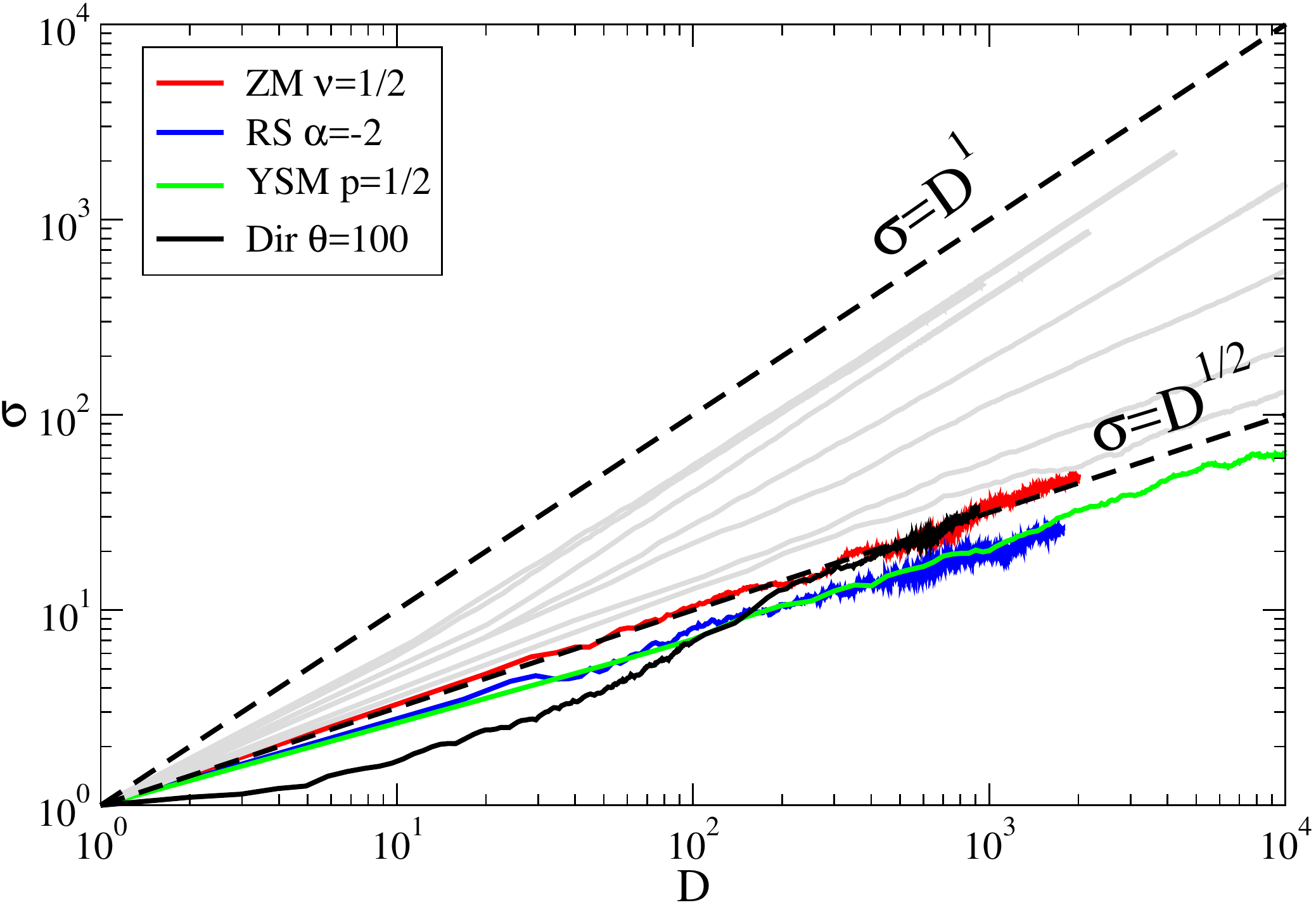}}
\caption{ 
	\textbf{Taylor's law in various generative models.}
	(Left panel)
    Models that do not display a square root dependence of the dictionary standard deviation versus the dictionary itself are shown in color, the others in gray. Curves are listed from top to bottom according to their visual ordering. The P\'olya's urn model with triggering (PUT) shows an exponent one when $\nu<\rho$ and exponents in the range from 1/2 to ca.\ 0.87 when $\nu \ge \rho$. The Poisson-Dirichlet (PD) process also displays a unity exponent.
	(Right panel)
    Models with a square root dependence of the dictionary standard deviation versus the dictionary itself are shown in color, the rest (highlighted in the left panel) in gray. The models are Zanette-Montemurro (ZM), Random Sampling (RS), Yule-Simon Model (YSM) and the Dirichlet process (Dir). All these four processes as well as the PUT with parameters $\nu=1$ and $\rho=2$, the PD with $\alpha=1/2$ and $\theta=1$ produce the same Heaps' law with exponent $\gamma=1/2$.
Each curve is the result of 100 runs of $10^6$ steps each. The dashed lines with exponents 1/2 and 1 are shown as a guide for the eye.
	\label{fig:taylor}}
\end{figure} 
\noindent
From Eqs.~(\ref{eqn1:heaps}-\ref{eqn3:heaps}) it is clear that randomly sampling a Zipf's law with a given exponent results in a Heaps' law with linear and sublinear exponents tuned by the exponent of the Zipf's. On the other hand, Eqs.~(\ref{eq:asymptotic_sub}-\ref{eq:asymptotic_log}) show that the PUT model is also producing the same Heaps' exponents with the same relation to the Zipf's exponent as in the random sampling. Therefore, one legitimate question is whether also the PUT is actually performing a kind of sophisticated random sampling of an underlying Zipf's law. One possible way to discriminate PUT from a random sampling is to look at the fluctuation scaling, i.e., the Taylor's law discussed in section~\ref{sec:IntroTaylor}, which connects the standard deviation $s$ of a random variable to its mean $\mu$. Simple analytic calculations \cite{gerlach2014} show that the Poissonian sampling of a power-law leads to a Taylor's law with exponent 1/2, i.e., $s\propto \sqrt\mu$. 

Real text analysis show instead a Taylor exponent of 1~ \cite{gerlach2014}, which points to the obvious conclusion that the process of writing texts is not an uncorrelated choice of words from a fixed distribution. In~\cite{gerlach2014} this was explained by a ``topic-dependent frequencies of individual words''. The empirical observation therein, was that the frequency of a given word changes according to the topic of the writing. For example, the term ``electron'' has an high frequency in physics books and low frequency in fairy tales, so that its rank is low in the first case and large in the second. The result is that there exist different Zipf's laws with the same exponent according to the topic and the enhanced variance of the dictionary size is ascribable to these multitude of Zipf's laws that add a further variability to the sampling process.

In PUT there is certainly no topicality as in real texts. Nevertheless, we find numerically a linear Taylor's law in case of sublinear Heaps' exponents ($\nu<\rho$). In PUT, there is no Zipf's law beforehand: it is built during the process instead and this is sufficient to boost the variance of the dictionary, on average, at any given time.

In Fig.~(\ref{fig:taylor}) we show the numerical results of two simulations of PUT with $\nu<\rho$, one with $\nu=\rho$ and one with $\nu>\rho$, in order to cover all the possible cases of Eqs.~(\ref{eq:asymptotic_sub}-\ref{eq:asymptotic_log}), plus the random sampling from a zipfian distribution with exponent $\alpha=-2$. Besides the interesting linearity of the fluctuation scaling in the case of $\nu<\rho$, also its behaviour in case of fast growing spaces $\nu>\rho$ can be pointed out. In that case Heaps' law is linear as shown in Eq.~\ref{eq:asymptotic_lin}, and the traditional model of reference is the Yule-Simon model (YSM)~\cite{Simon_1955}. The Yule-Simon model generates a sequence of characters with the following iterative rule. Starting from an initial character, at each time with a constant probability $p$ a brand new character is chosen while with probability $1-p$ one selects one of the characters already present in the sequence (which implies drawing them with their multiplicity). In this way, in YSM, the rate of growth of different characters is constant and equal to $p$ and this constant rate of innovation yields a linear Heaps' law. The preferential attachment rule leads to a Zipf's law with exponent $|\alpha|=1-p$. This is consistent with Eq.~(\ref{eqn3:heaps}) of random sampling and even with Eq.~(\ref{eq:asymptotic_lin}) of PUT. The difference between YSM and PUT can be appreciated with Taylor's law. In YSM, new characters appear with probability $p$ so that the average number of different characters at step $N$ is $\mu=pN$, and the variance $\sigma^2=Np(1-p)=\mu (1-p)$ as in the binomial distribution.  As a result, in YSM one gets the Poissonian result $\sigma\propto\sqrt\mu$. In contrast, PUT  features numerically an exponent of $\simeq 0.58$, i.e., larger than $1/2$ but still less than 1 (see Fig.~\ref{fig:taylor}).

\noindent Given the intrinsic inability of YSM to accomplish for sub-linear dictionary growths, Zanette and Montemurro~\cite{zanette2005} proposed a simple variant of it. In this variant (ZM), the rate of introduction of new characters, i.e., $p$, is not constant anymore. It is made instead time-dependent with an \emph{ad hoc} chosen functional form able to reproduce the right range for the Heaps' exponents. For a Heaps' exponent $\gamma$, the rate of innovation $p$ is chosen  proportional to $t^{\gamma-1}$. This expedient allows to reproduce both Zipf's law and, by construction, Heaps' law. The two mechanisms for Zipf's and Heaps' production are independent from each other as in YSM so that we expect for Taylor's law the same behavior of YSM, i.e., an exponent $0.5$.  After all, ZM can be seen as a YSM with a diluting time flow, which might not affect the scaling of the fluctuations of YSM at a given time. In Fig.~\ref{fig:taylor} we show that indeed ZM features a Taylor's exponent of $0.5$ (magenta curve).

For the Poisson-Dirichlet  and the Dirichlet processes, analytical solutions can be computed for the moments of the probability distribution $P(D(n))$~\cite{Yamato2000, buntine2010}, yielding the asymptotic exponents  respectively $1$ and $1/2$ in the Taylor's law. Numerical results are given in Fig.~\ref{fig:taylor}. Note that a non trivial exponent in the Taylor's law is featured by the  Poisson-Dirichlet   process, where the probability of a novelty to occur does depend on the number of previous novelties, while the Dirichlet   process lacks both properties.

\section{Discussion}
\noindent
In this paper we have argued that the notion of adjacent possible is key to explain the occurrence of the Zipf's, Heaps' and Taylor's laws  in a very general way. We have presented a mathematical framework, based on the notion of adjacent possible, and instantiated through a Polya's urn modelling scheme, that accounts for the simultaneous validity of the three laws just mentioned in all their possible regimes. 

We think this a very important result that will help in assessing the relevance and the scope of the many approaches proposed so far in the literature. In order to be as clear as possible let us itemize the key points:

\begin{itemize}

\item The first point we make is about the many claims made in literature about the possibility to deduce the Heaps' law by simply sampling a Zipf-like distribution of frequencies of events. Though, as seen above,  it is possible to deduce a power-law behaviour for the growth of distinct elements by randomly drawing from a Zipf-like distribution, this procedure does not allow to reproduce the empirical results. It has been conjectured in~\cite{gerlach2014}, that texts are subject to a topicality phenomenon, i.e., writers do not sample the same Zipf's law. This implies that the same word can appear at different ranking positions depending on the specific context. Although this is an interesting point, we think that the deduction of the Heaps' law from the sampling of a Zipfian distribution is not satisfactory from two different points of view.  First of all  the empirical Heaps' and Zipf's laws are never pure power-laws. We have seen for instance that for written texts the frequency-rank plot features a double slope. Nevertheless we have seen that a relation exists between the exponent of the frequency-rank distribution at high ranks (rare words) and the asymptotic exponent of the Heaps' law. In other words, the behaviour of the rarest words is responsible for the entrance rate of new words (or new items). Even though a pure power-law behaviour was observed, we have shown that the statistics of fluctuations, represented by the Taylor's law, would not reproduce the empirical results (unless a specific sampling procedure based on the hypothesis of topicality is adopted~\cite{gerlach2014}). The conclusion to be taken is that in general the Heaps' and the Zipf's laws are non-trivially related and their explanation should be made based instead on first-principle. 

\item Models featuring a fixed space of possibilities are not able to reproduce the simultaneous occurrence of the three laws. For instance a multicolor Polya's urn model~\cite{polya_multicolor_1997} does not even produce power-law-like behaviours for the Zipf's and the Heaps' laws. It rather features a saturation phenomenon, related to the exploration of the predefined boundaries of the space of possibilities. The conclusion here is that one needs a modelling scheme featuring a space of possibilities with dynamical boundaries, for instance expanding ones. 

\item Models that incorporate the possibility to expand the space of possibilities like the Yule-Simon~\cite{Simon_1955} model or the Hoppe model fail in explaining the empirical results. In the Yule-Simon model the innovation rate is constant and the the Heaps' law is reproduced with the trivial unitary exponent. An ad-hoc correction to this has been proposed by Zanette and Montemurro~\cite{zanette2005}, who postulate a sublinear power-law Heaps's law form the outset, without providing any first-principle explanation for it. Also in this case the result is not satisfactory because the resulting time-series does not obey the Taylor's law, being instead compatible with a series of i.i.d variables. The question is now why this approach is not reproducing Taylor's law despite the fact that it fixes the expansion of the space of possibilities. In our opinion what is lacking in the scheme by Zanette and Montemurro is the interplay between the preferential attachment mechanism and the exploration of new possibilities. In other words, the triggering effect which is instead a key features of the PUT model (see next item). Different is the situation for the Hoppe model~\cite{Hoppe_1984}, i.e., a multicolor Polya's urn with a special replicator color. In this case, though a self-consistent expansion of the space of possibilities is in place, an explicit mechanism of triggering, in which the realization of an innovation facilitates the realization of further innovations, lacks. In this case the innovation rate is too weak and the Heap's law features only a logarithmic growth, i.e., slower than any power-law sublinear behaviour.

\item The Polya's urn model with triggering (PUT)~\cite{adjacent_possible_2014}, incorporating the notion of adjacent possible, allows to simultaneously account for the three laws, Zipf's, Heaps' and Taylor's, in all their regimes, without ad-hoc or arbitrary assumptions. In this case the space of possibilities expands conditional to the occurrence of novel events in  a way that is compatible with the empirical findings.  From the mathematical point of view, the expansion into the adjacent possible solves another issue related to Zipf's and Heaps' generative models. In fact, in PUT one can switch with continuity from the sublinear to the linear regime of the dictionary growth and vice-versa and this by tuning one parameter only: the ratio $\nu/\rho$. This ratio is not limited to a ratio of integers. In fact, in the SI of~\cite{adjacent_possible_2014} it was demonstrated that the same expressions for the Heaps' and Zipf's  laws are recovered if one uses parameters $\rho$ and $\nu$ extracted from a distribution with fixed means. One possible strategy is to fix an integer $\rho$ while $\nu$ can assume any value in the real numbers (in simulations this is a floating point value), and the mantissa can be taken into account by resorting to probabilities. Therefore, it is perfectly sound to state that one switches \emph{with continuity} from the sublinear regime to the linear one in the interval $|\nu/\rho-1|<\varepsilon$, with $\varepsilon\ll1$, although the rigorous mathematical characterization of the transition is far from being understood.

\item It should be remarked that  the Poisson-Dirichlet process~\cite{csp_pitman2006,buntine2010, deblasi2015} is also able to explain the three Zipf's, Heaps' and Taylor's laws only in the strict sub-linear regime for the Heaps' law. It cannot however account for a constant innovation rate as in the PUT modelling scheme. We also point out that the PUT model embraces the Poisson-Dirichlet and the Dirichlet processes as particular cases. 

\end{itemize}

In this paper we highlighted that the simultaneous occurrence of the Zipf's, Heaps' and Taylor's laws can be explained in the framework of the adjacent possible scheme. This implies considering a space of possibilities that expands or gets restructured conditional to the occurrence of a novel event. The P\'olya's urn with triggering features these properties. Poisson-Dirichlet processes also can be said belonging to the adjacent possible scheme. Though no explicit mention is made about the space of possibilities in those schemes, the probability of the occurrence of a novel events closely depends on how many novelties occurred in the past. We recall that the PUT model includes Dirichlet-like processes as particular cases. From this perspective PUT-like models seem to be good candidates to explain higher order features connected to innovation processes. We conclude by saying that the very notion of adjacent possible, though sufficient to explain the stylized facts of innovation processes, can be only conjecture also as a necessary condition for the validity of the three laws mentioned above.  No counterexamples have been found so far, in fact, that, without a dynamically restructured space of possibilities, one can satisfactorily explain the empirically observed laws.

\vspace{1cm}
\noindent \textbf{Acknowledgments:}\\
We thank S.~H.~Strogatz for interesting discussions.
VDPS is grateful to M.~Osella for interesting discussions on the relevance of Taylor's law.\\
VDPS acknowledges the Austrian Research Promotion Agency FFG under grant $\#$857136 for financial support.


\vspace{0.5cm}

\noindent \textbf{The following abbreviations are used in this manuscript:}\\[3mm]
\begin{tabular}{rl}

\textbf{PUT}: & P\'olya's Urn with Triggering \cite{adjacent_possible_2014}\\
\textbf{PD}: & Poisson-Dirichlet process\\
\textbf{Dir}: & Dirichlet process\\
\textbf{RS}: & Random Sampling\\
\textbf{YSM}: & Yule-Simon Model \cite{Simon_1955}\\
\textbf{ZM}: & Zanette-Montemurro model \cite{zanette2005}
\end{tabular}

\begin{appendices}
\section{Analytic derivation of Heaps' law in the urn model with triggering}
\label{sec:appendix_A}
\noindent
We here derive the Heaps' law for a general variation of the urn model with triggering. For the sake of completeness, we recall here the  model:
An
urn $\mathcal{U}$ initially contains $N_0$ distinct elements,
represented by balls of different colors. By randomly extracting elements from the urn, we construct a 
sequence $\mathcal{S}$.
 Both the urn and the sequence enlarge
during the process. At each time step $t$, an element $s_t$ is drawn at random from the urn:
(i)  iff the chosen element $s_t$ is old (i.e., it already appeared in the sequence $\mathcal{S}$), it is added to
  the sequence, and put back in the urn along with $\rho$
  additional copies of it;
  (ii) iff the chosen element $s_t$ is new (i.e., it appears for
  the {\em first time} in the sequence $\mathcal{S}$), 
  it is added to
  the sequence, and put back in the urn along with $\tilde{\rho}$
  additional copies of it. Further,
  $\nu+1$
  brand new distinct elements are also added to the urn.
 
 We can now write the equation governing the growth of the number of distinct elements $D(n)$  as a function of the total number $n$ of elements in the sequence ($n$ is also obviously denoting the time step $t$ above):

\begin{equation}
  \frac{d D}{dn}= \frac{N_0 + \nu D}{N_0+ a D+\rho n},
\end{equation}
where we have defined $a=\nu+1-\rho+\tilde{\rho}$.

By defining $z=\frac{D}{n}$ and  neglecting $N_0$, we can write:
\begin{equation}
  \frac{d z}{dn}= \frac{1}{n} \frac{d D}{n} - \frac{D}{n^2} \,\,\, \Rightarrow \,\,\, \frac{d D}{dn}=  n \frac{d z}{d n} + z = \frac{ \nu z}{ a z+\rho }
  \end{equation}
which gives:  
\begin{equation}\label{eq:integral}
\int_{z(n_0)}^{z(n)}  \frac{ a z+\rho }{ z( \nu -a z -\rho)} d z= \int_{n_0}^{n}\frac{ dn }{ n } .
  \end{equation}
Here  we note that by definition $0\leq z \leq 1$, and   $z(n_0)=D(n_0)/n_0$, for a given $n_0$ such that the solutions we found are valid for any $n\geq n_0$. In order to integrate equation~\ref{eq:integral} we need to study the sign of the expression  $\nu -a z -\rho$. Let us do this by considering separately the case $\rho>\nu$ and $\rho<\nu$, and postponing the computation for $\rho=\nu$.

\subsection{Case 1: $\rho>\nu$}

In this case, if $a \geq 0$ we have $\nu -a z -\rho < 0$  (and thus obviously $a z +\rho -\nu > 0$), while if $a<0$   it exists a $z_0$ such that  $\nu -a z -\rho < 0$  for $z<z_0$. Thus, if $z(n)$ is decreasing in $n$, we can safely perform the integration for any $n\geq n_0$, for some $n_0$. Let us make this assumption and verify it at the end of the computation.
By integrating equation~\ref{eq:integral} we thus obtain:
\begin{equation}
-\log{( a z +\rho - \nu)}  (1+\frac{\rho}{\nu-\rho})  \mid_{z(n_0)}^{z(n)} + \frac{\rho}{\nu-\rho} \log{z} \mid_{z(n_0)}^{z(n)} = \log{n} \mid_{n_0}^{n}
  \end{equation}
and solving:
\begin{equation}
(a z +\rho - \nu)^\nu= A n^{\rho-\nu} z^\rho  ~,\,~~ A=\exp(C) ~,\,~~ C = \log{\frac{(a z(n_0) +\rho - \nu)^\nu}{z(n_0)^\rho}} - \log{n_0^{\rho-\nu}} .
  \end{equation}
We can now substitute $z=\frac{D}{n}$, and after some algebra we can write:
\begin{equation}
  D - \frac{a}{ A^{\frac{1}{\nu}} } D^\frac{\nu}{\rho}  = B  n^\frac{\nu}{\rho} ~,\,~~ B=A^{-^{\frac{1}{\rho}} } (\rho - \nu)^\frac{\nu}{\rho}
 \end{equation}
that gives the solution:
\begin{equation}\label{eq:D_caso1}
  D(n) = B  n^\frac{\nu}{\rho} +O(n^\frac{\nu^2}{\rho^2}).
 \end{equation}
We observe that $\lim_{n \to \infty} z(n)= \lim_{n \to \infty} \frac{D(n)}{n} = 0$ monotonically, so that the assumption made above is satisfied.

\subsection{Case 2: $\rho<\nu$}
\noindent
Let us now assume $\nu -a z -\rho > 0$ and let us verify the assumption at the end.
We thus write:
\begin{equation}
-\log{( \nu -a z -\rho )}  (1+\frac{\rho}{\nu-\rho})  \mid_{z(n_0)}^{z(n)} + \frac{\rho}{\nu-\rho} \log{z} \mid_{z(n_0)}^{z(n)} = \log{n} \mid_{n_0}^{n} .
  \end{equation}
After similar calculation as in the case $\rho>\nu$, we arrive to the relation:
\begin{equation}
   D + \frac{A^\frac{1}{\nu}}{ a} D^\frac{\rho}{\nu}  = \frac{\nu-\rho}{a}  n ~,\,~~ A=\exp(C)   ~,\,~~ C = \log{\frac{(\nu-a z(n_0) -\rho )^\nu}{z(n_0)^\rho}} - \log{n_0^{\rho-\nu}}
   \end{equation}
that gives the solution:
\begin{equation}\label{eq:D_caso2}
  D(n) = \frac{\nu-\rho}{a}  n -   \frac{A^\frac{1}{\nu}}{ a} O(n^\frac{\rho}{\nu})
 \end{equation}
Note that $a\geq 0$ per $\nu>\rho$. We already discussed in the main text the case $a=0$, that has to be treated separately, so we here consider $a>0$. From~\ref{eq:D_caso2} we observe that 
 $z(n)=D(n)/n$ is increasing  in $n$:   $z(n)=  \frac{\nu-\rho}{a} - \tilde{z}(n)$ with $\lim_{n \to \infty} \tilde{z}(n)=0$. Thus 
$\nu -\rho-a z  = a \tilde{z}(n) > 0$, with limit zero in the asymptotic limit $n \to \infty$. The initial assumption is thus satisfied in the entire range of the $z$ values. 

\section{Analytic determination of Zipf's law in the continuous approximation}
\label{sec:appendixB}

\noindent
In the following we derive the expression of Zipf's exponents for the model of P\'olya's urn with innovations, by exploiting the  continuous approximation.
\subsection{Preliminary considerations}
The time evolution of the number of different colors $D$ in the stream can be approximated by the following equation (see Eq.~(\ref{eq:D_complete_text}) with $N_0=1$ ):
 \beq
 	\dot D = \frac{1+\nu D}{1+\rho t + (\nu+1) D}~~~\mbox{with}~~~D(0)=1.
 	\label{eq:dictionary}
\eeq
Putting aside the particular cases $\nu=0$ and $\nu=\rho$, Eq.~(\ref{eq:dictionary}) can be solved analytically to yield, in the leading terms at large $t$ (see also Appendix~\ref{sec:appendix_A}) :
\beq
D(t) \approx \left\{ 
\begin{array}{cr}
	\frac{\nu-\rho}{\nu+1}\,t&~~~\mbox{if}~~~\nu>\rho\\
	\left(\frac{\rho(\rho-\nu)}{\nu(\rho-1)+2\rho}\,t\right)^{\nu/\rho}&~~~\mbox{if}~~~\nu<\rho
\end{array}
\right.
~~~\mbox{with}~~~ t\gg1.
\eeq
The two regimes given by the relative values of $\nu$ and $\rho$ result in two different Heaps' exponents $\gamma$, i.e., $\gamma=1$ and $\gamma=\nu/\rho$.

In the denominator of Eq.~(\ref{eq:dictionary}), the total number of balls in the urn appears: 
\(N(t) = 1+\rho t + (\nu+1) D\), 
so that we can write:
\beq
N(t) = \frac{\nu D}{\dot D} \approx \frac{\nu t}{\gamma} \approx
\left\{ 
\begin{array}{l}
	\nu t~~~\mbox{if}~~~\nu>\rho\\
	\rho t~~~\mbox{if}~~~\nu<\rho
\end{array}\right.
.
\eeq
\subsection{Master equation}
We denote with $N_k$ the number of balls with a given color occurring $k$-times in the urn.
In particular we have $\sum_k N_k = D$.
The following master equation can be written for the $N_k$:
\beq
\frac{\partial N_k}{\partial t}= \frac{(k-\rho) N_{k-\rho}}{N(t)}-\frac{k N_{k}}{N(t)} \approx
	-\frac{\rho}{N(t)}\frac{\partial kN_k}{\partial k}\approx
	-\frac{\rho\dot D}{\nu D}\frac{\partial kN_k}{\partial k}.
	\label{eq:cont_eq}
\eeq
We introduce now the probability $p_k$ that a given color appears $k$-times in the urn, i.e., the corresponding normalized version of the number of occurrences $N_k$. 
In order to have $\sum p_k =1$, we must choose $N_k=Dp_k$.
The idea is that, as the time runs, the probabilities $p_k$ will tend to a stationary distribution, i.e., a distribution independent of $t$.
By substituting $N_k=Dp_k$ in Eq.~(\ref{eq:cont_eq}), we get
\beq
p_k= -\frac{\rho}{\nu}\frac{\partial kp_k}{\partial k}.
\label{eq:final_pk}
\eeq
This equation can be solved easily by substituting $p_k\propto k^{-\beta}$ and solving for $\beta$, which leads to
\beq
\beta = 1+\frac{\nu}{\rho}
\eeq
and conversely the frequency-rank exponent $\alpha=\rho/\nu$.
Note that, while $\gamma$ depends on the relative values of $\nu$ and $\rho$, $\alpha$ does not.
To relate the $p_k$ distribution in the urn to that of the stream is an easy task.

\subsection{Particular case $\nu=\rho$}
Also in this case Eq.~(\ref{eq:dictionary}) can be solved analytically with the solution including the Lambert $W$ function.
At large values of $t$, the solution can be approximated as
\beq
	D(t)\approx \frac{\nu}{\nu+1}\,\frac{t}{\log t},
\eeq
so that $N(t)\approx \nu t$.
Eq.~(\ref{eq:cont_eq}) can be written as before as:
\beq
\frac{\partial D p_k}{\partial t}= 
	-\frac{\nu\dot D}{\nu D}\frac{\partial D k p_k}{\partial k}\approx
	~~~\longrightarrow~~~ p_k=-\frac{\partial kp_k}{\partial k},
\eeq
which results in $\beta=2$ and $\alpha=1$. 

\subsection{Particular case $\nu=0$}
This case is identical to the Hoppe's urn model.
When a ball with a brand new color is extracted, exactly one new color enters the urn so that the number of unobserved colors stays the same during the whole dynamics.
If we start with one single ball, there will always be only one unobserved color in the urn and this color would have exactly the same function of the black ball with weight one in Hoppe's model.
The equation for the growth of novelties will be:
\beq
	\dot D = \frac{1}{1+\rho t + D}~\stackrel{t\rightarrow\infty}{\approx}~\frac{1}{\rho t}
	~~~\longrightarrow~~~ D\approx \frac{1}{\rho}\log t,
\eeq
while the frequency-rank will be decaying exponentially.
In order to introduce the equivalent counterpart of the weight of the black ball in the Hoppe's model, whenever a novelty is extracted, $w$ balls of the same brand new color could be added to the urn.

\end{appendices}


\end{document}